# An analysis of the technology acceptance model in understanding university students' behavioral intention to use metaverse technologies

Nikolaos Misirlis[1], Harris Bin Munawar[1]

1 HAN University of Applied Sciences, International School of Business, The Netherlands

**Abstract**

*The last trend in technology is the upcoming Metaverse [1]. Metaverse represents a combination of virtual and augmented technology. With this technology, users will be able to immerse into a fully digital environment by obtaining a virtual identity through a digital avatar and acting as this was the real world. They can meet other users, shop, buy real estate, visit bars and restaurants, even flirt. Metaverse can be applied in several aspects of life such as (among others): Economy (Metaverse is entering into the cryptocurrency field), finance [2], social life, working environment, healthcare, real estate [3], and education [4]. In the last 2 and a half years, during the COVID-19 pandemic, universities made immediate use of e-learning technologies, providing students with access to online learning content and platforms. Previous considerations on how to better integrate the technology to universities or how the institutions can be better prepared in terms of infrastructures were vanished almost immediately due to the necessity of immediate actions towards the need for social distance and global health [5]. The present study proposes a framework for university students' metaverse technologies in education acceptance and intention to use. The study is based on the Technology Acceptance Model (TAM) [6, 7]. The objectives of the study are to analyze the relationship of students' intention to use metaverse in education technologies (hereafter named MetaEducation) in correlation with selected constructs of TAM such as: Attitude (ATT), Perceived Usefulness (PU), Perceived Ease of Use (PE), Self-efficacy (SE) of the metaverse technologies in education, and Subjective Norm (SN). The present study develops a structural model of MetaEducation acceptance. This model will be useful to universities' managers, policymakers and professors to better incorporate the upcoming metaverse technology. The present study tests (if supported) the correlations among the aforementioned constructs. Preliminary results show a hesitance to use MetaEducation technologies from university students. Self-efficacy and Subjective Norm affect Attitude and Perceived Usefulness positively, but on the other side, there is no strong correlation between Perceived Ease of Use and Attitude or Perceived Usefulness and Attitude. Authors believe that the weak ties among the studies constructs have to do with the lack of knowledge of what really MetaEducation really is, and which are its advantages of use.*

## 1. Introduction

The concept of virtual reality is not new. Already 25 years ago, the first SCI-FI movies introduced to the audience concepts related to new/future technologies. The first time when the term Metaverse was introduced, was back in 1992 in the SCI-FI novel 'Snow Crash'. Metaverse is a combination of virtual and augmented reality on a 30-70% balance [1]. Users, after picking an avatar, are able to visit digital worlds, interacting with other users. These digital worlds include digital cities, concerts, night-life, even schools. It seems that the fields in that Metaverse can be applied are endless. From economy to finance [2], to social life and work, eHealth [8] or even real estate [3], the possibilities keep on rising. Of course, education is one of those fields [4].

Regarding education, after more than two years of the Pandemic, almost all schools incorporated digital facilities for students [9]. Most of them will keep the hybrid education even after the end of the Pandemic. Universities, before the Pandemic, were sceptical of the use of MetaEducation, but today a new hybrid environment is more than a necessity for schools.

## 2. Research Objectives
The present study proposes a theoretical framework, based on the Technology of Acceptance Model, of tertiary students' acceptance of metaverse technologies in education. The study analyzes students' behavioral intention toward the use of the aforementioned technology. The objective of the study was to analyze the relationship and the correlations of students' intention to use metaverse technologies in education with specific components such as attitude, perceived usefulness, perceived ease of use, self-efficacy, and subjective norm. The linear structural model created will provide university academics, managers and policymakers the knowledge for better implementing the structures (new modules, technology needed, ethics committee, policies, data protection rules, etc.) and adapting the curricula for the future.

## 3. Research Hypotheses
The study will test and support the following hypotheses:
H1: University students' behavioral intention to use MetaEducation is affected by: Attitude, Perceived Usefulness, Perceived Ease of Use, Self-efficacy, or Subjective Norms.
H2: University students' attitude toward the use of MetaEducation is affected by: Perceived Usefulness, Perceived Ease of Use, Self-efficacy, or Subjective Norms.
H3: University students' Perceived Usefulness to use MetaEducation is affected by: Perceived Ease of Use, Self-efficacy, or Subjective Norms.
H4: University students' Perceived Ease of Use towards the use of MetaEducation is affected by: Self-efficacy, or Subjective Norms.

## 3. Literature review
The present research will use the well-known Technological Acceptance Model (TAM), first introduced by Davis [10] as an extension of Ajzen and Fishbein's Theory of reasoned action - TRA [11]. As aforementioned, TAM is using Subjective Norms, Perceived Usefulness, Perceived Ease of Use, Self-efficacy, and attitude as control variables, dependent and independent to predict the behavioral intention of users towards the use of technology and its acceptance. Several studies focusing on education are already conducted [12-14] , but to the best of our knowledge, this is the first one searching the acceptance behavior of technologies applied in education, related to the Metaverse.

## 4. Research design and methodology
Based on previous research, a theoretical model was developed. Figure 1 represents the model to be tested and analyzed.

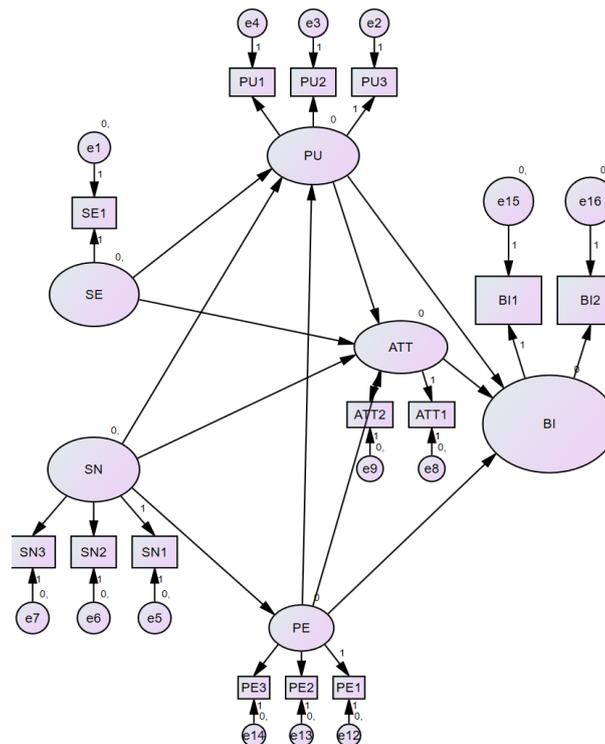

Fig. 1: Theoretical model

The directed arrows show the relationship between the latent variables and the observed ones. PE and PU can be considered cognitive constructs. Based on the above theoretical model, the dataset of our survey was applied to produce the measurement model (Figure 2). The population of the valid respondents consists of 285 international university students at HAN University of Applied Sciences in The Netherlands. A size of 115 respondents would be an appropriate minimum sample size for SPSS Amos to produce valid results. 63% are defined as males and 34% as females. The percentages on faculties and year of study are equally distributed. The dominant nationality is the Dutch (47%) and the rest 53% is equally distributed among 35 different nationalities. The data were first recorded in MS Excel and later transferred to SPSS, analyzed with Amos.

## 5. Analysis of the measurement model

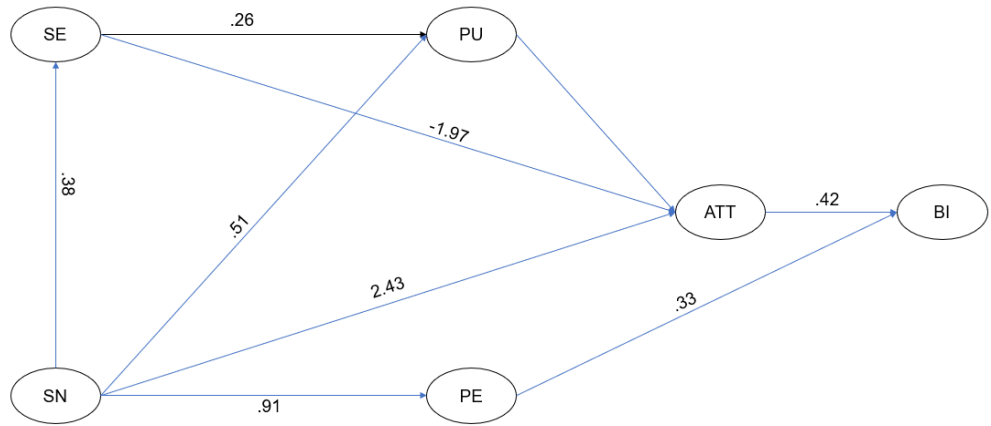

Fig. 2: Structural model

Self-efficacy is negatively correlated to attitude, but on the contrary, subjective norms are positively correlated. There is no strong correlation between Perceived Ease of Use and Attitude or Perceived Usefulness and Attitude. The overall model is still weak, even if it is statistically acceptable. Table 1 summarizes (in bold) the findings that can lead to positive or negative correlations. From the same table, we notice that the final behavior of students is slightly affected by the Perceived ease of use but not by the perceived usefulness.

Table 1: Total effects of the model

|     | SN    | SE     | PU   | PE   | ATT  | BI   |
|-----|-------|--------|------|------|------|------|
| PE  | **.914**  | .000   | .000 | .000 | .000 | .000 |
| ATT | **2.425** | **-1.974** | .217 | .000 | .000 | .000 |
| BI  | **1.318** | **-.829**  | .091 | .328 | .420 | .000 |
| SE1 | .000  | **1.000**  | .000 | .000 | .000 | .000 |

An explanation of those results, based on our experience in Academia, is related to the fact that students are not yet fully aware of the Metaverse technology or its possible applications in tertiary education. A 55.4% of the students believe that their social life depends on technology, in general (n:158/285), but on the other side, 60.4% (n:171) and 50% (n:142) disagree that their personal happiness and their well-being, respectively, are somehow depending from the technology. This final result is contradictory to the next result, showing that 66.1% (n:187) of the students switch on digital to unwind. Regarding the use of MetaEducation technologies, 76.7% (n:217) believe that education is dependent from that technology, even if there are in general, still skeptical toward Metaverse and MetaEducation. Regarding the associations of Metaverse and the relationship with family and friends, the results show an equal distribution between those who believe that technology can strength relationships (41%, n:117) and those who disagree (39.6%, n: 112).